\documentclass{PoS}
\PoS{PoS(HEP2005)258}				     
\newcommand{\bhh}{$B^0_{(s)}\rightarrow h^{+}h^{'-}$}
\newcommand{\bshh}{$B^0_{s}\rightarrow h^{+}h^{'-}$}
\newcommand{\bdhh}{$B^0 \rightarrow h^{+}h^{'-}$}
\newcommand{\bdkpi}{$B^0\rightarrow K^{+} \pi^{-}$}
\newcommand{\bdpipi}{$B^0\rightarrow \pi^{+} \pi^{-}$}
\newcommand{\bskk}{$B^0_{s}\rightarrow K^{+} K^{-}$}
\newcommand{\bskpi}{$B^0_{s}\rightarrow K^{-} \pi^{+}$}
\newcommand{\bspipi}{$B^0_{s}\rightarrow \pi^{+} \pi^{-}$}
\newcommand{\bdkk}{$B^0\rightarrow K^{+} K^{-}$}
\newcommand{\bd}{$B^0$}
\newcommand{\bs}{$B^0_s$}

\newcommand{\intlumipb}{pb$^{-1}$}

\newcommand{\massmev}{MeV/$c^2$}

\usepackage[abs]{overpic}
\title{Branching fractions of \bhh\ modes at CDF}
\ShortTitle{Branching fractions of \bhh\ modes at CDF}

\author{\speaker{Diego Tonelli}\thanks{representing the CDF Collaboration.}\\
        Istituto Nazionale di Fisica Nucleare - Sezione di Pisa\\
	Edificio C - Polo Fibonacci - Largo B.~Pontecorvo 3, Pisa, Italy\\
        E-mail: \email{diego.tonelli@pi.infn.it}}

\abstract{
I report the analysis of \bhh\ decays  (where $h$ and $h^{'}$ denote $K$ or $\pi$),
 in  180 \intlumipb~ of proton-antiproton collisions at $\sqrt{s}= 1.96$ TeV, with the CDF II detector at the Tevatron Collider.\\
A \bhh\ signal was reconstructed at a hadron collider for the first time, in the sample
 collected by the dedicated trigger on long-lived decays. With  
about 900 signal events, we observed  the new mode \bskk,
and we measured its branching fraction relative to the \bdkpi\ mode:
\begin{eqnarray*}
\frac{f_s}{f_d}\cdot\frac{\mathcal{B}(B^0_s \rightarrow K^+ K^-)}{\mathcal{B}(B^0 \rightarrow K^+\pi^-)}= 0.46 \pm 0.08~\mathit{(stat.)} \pm 0.07~\mathit{(syst.)};
\end{eqnarray*}
we also measured the \textsf{CP}-violating decay-rate asymmetry for the \bdkpi\ mode:
\begin{eqnarray*} 
A_{\mathsf{CP}}(B^0 \rightarrow K^+ \pi^-)= \frac{N(\bar{B}^0 \rightarrow K^-\pi^+)- N(B^0 \rightarrow K^+\pi^-)}{N(\bar{B}^0 \rightarrow K^-\pi^+)+N(B^0 \rightarrow K^+\pi^-)}=-0.013\pm 0.078~\mathit{(stat.)} \pm 0.012~\mathit{(syst.)}.
\end{eqnarray*}
Since we did not find evidence for rarer modes,
 we set the following 90\% C.L. upper limits on the decay rates:
\begin{eqnarray*}
  \frac{f_s}{f_d}\cdot\frac{\mathcal{B}(B^0_s \rightarrow K^- \pi^+)}{\mathcal{B}(B^0 \rightarrow K^+\pi^-)}< 0.08 ~~~~\mathrm{and}~~
\frac{\mathcal{B}(B^0_s\rightarrow \pi^+\pi^-)}{\mathcal{B}(B^0_s \rightarrow K^+ K^-)}< 0.05\nonumber
 \end{eqnarray*}
that are greatly improved with respect to current world averages.}
\FullConference{International Europhysics Conference on High Energy Physics\\ 
July 21st - 27th 2005\\
Lisboa, Portugal}
\begin{document}
\section{Introduction}
\label{sec:intro}
Today, the $b$-sector of flavor dynamics is being widely investigated in searching for
anomalous \textsf{CP}-violation effects. Direct measurements of the quark-mixing parameters from $b$-hadrons are compared with indirect
 results from global fits, related to other sectors, to look for deviations from 
the Standard Model  predictions. This approach may not be  straightforward, since the interpretation of 
 $b$-physics results is often affected by large uncertainties coming
from non-perturbative QCD effects: hence the need to simplify the problem by invoking
 symmetries under which the unknowns (partially) cancel. In this respect,
the joint study of two-body decays of \bd\ and \bs\ mesons into light, charged, pseudo-scalar mesons
 ($K^+K^-$, $\pi^+\pi^-$ and $K^+\pi^-$)\footnote{Unless otherwise stated, \textsf{CP}-conjugated
modes are implied,  and all branching fractions indicate \textsf{CP}-averages.} plays a key role, since these modes are related
 by subgroups of SU(3) symmetry \cite{fleischer}.  
CDF at the Fermilab Tevatron $p\bar{p}$ collider is the first and only experiment that has
 simultaneous access to \bd\ and \bshh\ decays. The CDF II detector is a multipurpose
magnetic spectrometer, surrounded by $\sim 4\pi$ calorimetry and muon chambers. Its components relevant for
 this analysis are the 
tracking system, composed by a silicon micro-vertex and 
an outer drift chamber, and the trigger; for further detector  details, see \cite{tdr}. 

\section{Trigger and dataset}
\label{sec:trigger}          
 The study of \bhh\ decays at a hadron collider is challenging, since one searches 
for generic final states in huge QCD backgrounds;  high ($\sim$ 50/event) track-multiplicities
and tiny signal-to-noise ratios at production, $(\sigma_{b\bar{b}}/\sigma_{p\bar{p}})\cdot\mathcal{B}\sim 10^{-9}$, make crucial the role of the trigger.
The Silicon Vertex Trigger (SVT) \cite{SVT}  exploits 
the $\sim$1.5 ps lifetime of $b$-hadrons, which, combined with their boost, results in decay vertices 
separated by hundreds of microns from  the $p\bar{p}$ vertices. 
%The SVT finds, in the plane transverse to the beam, silicon tracks in the level-2 trigger
 %with 25 $\mu$s latency, and  identifies displaced  tracks with offline-quality resolution\footnote{The SVT resolution combined 
%with the beam-width determines the total resolution  $\sigma_{SVT}\oplus\sigma_{beam}\simeq 47$ $\mu$m.} on their impact
% parameters\footnote{The i. p. ($d_0$) is
% the distance between the track's momentum and the
% $p\bar{p}$ vertex in the plane transverse to the beam.}.\par 
At the level-2 of the trigger, with 25 $\mu$s latency, the 
SVT finds silicon tracks in the plane transverse to 
the beam, and  identifies displaced tracks 
with offline-quality resolution\footnote{The SVT resolution combined 
with the beam-width determines the total resolution  $\sigma_{\mathit{SVT}}\oplus\sigma_{beam}\simeq 47$ $\mu$m.} on their impact
 parameters\footnote{The i. p. ($d_0$) is
 the distance between the track's momentum and the
 $p\bar{p}$ vertex in the plane transverse to the beam.}.\par 
We analyzed a $\approx$ 180 \intlumipb~ sample of 
 pairs of oppositely-charged tracks with $p_T > 2$ GeV/$c$,
  $p_T(1) + p_T(2) > 5.5$ GeV/$c$, and a transverse
 opening-angle $20^\circ < \Delta\phi < 135^\circ$. We required each track to have
an impact parameter 100 $\mu$m $< d_0 < 1$ mm.
 The $B$ meson candidate was required to have an impact parameter $d_0(B)< 140~\mu$m 
 and a transverse decay-length $L_{xy}(B)>200~\mu$m.
\section{Signal extraction}
\label{sec:signa_extraction}
\begin{figure}[!ht] 
\begin{center}
\begin{overpic}[width=4.69cm,height=4.69cm,grid=false,tics=1]{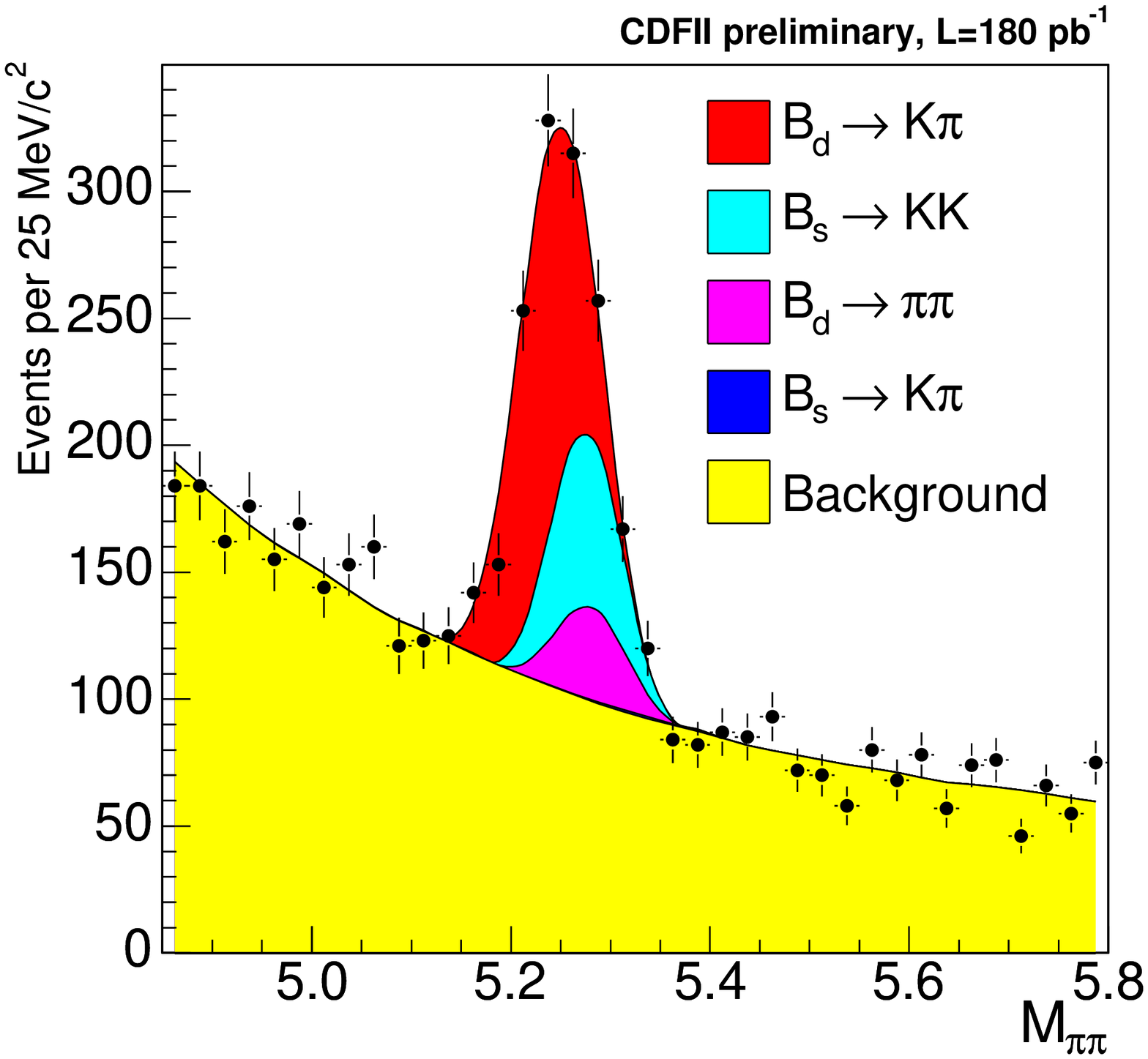}
\put(8,39){\scriptsize{\emph{(a)}}}								
\end{overpic}   
\begin{overpic}[width=4.69cm,height=4.69cm,grid=false,tics=1]{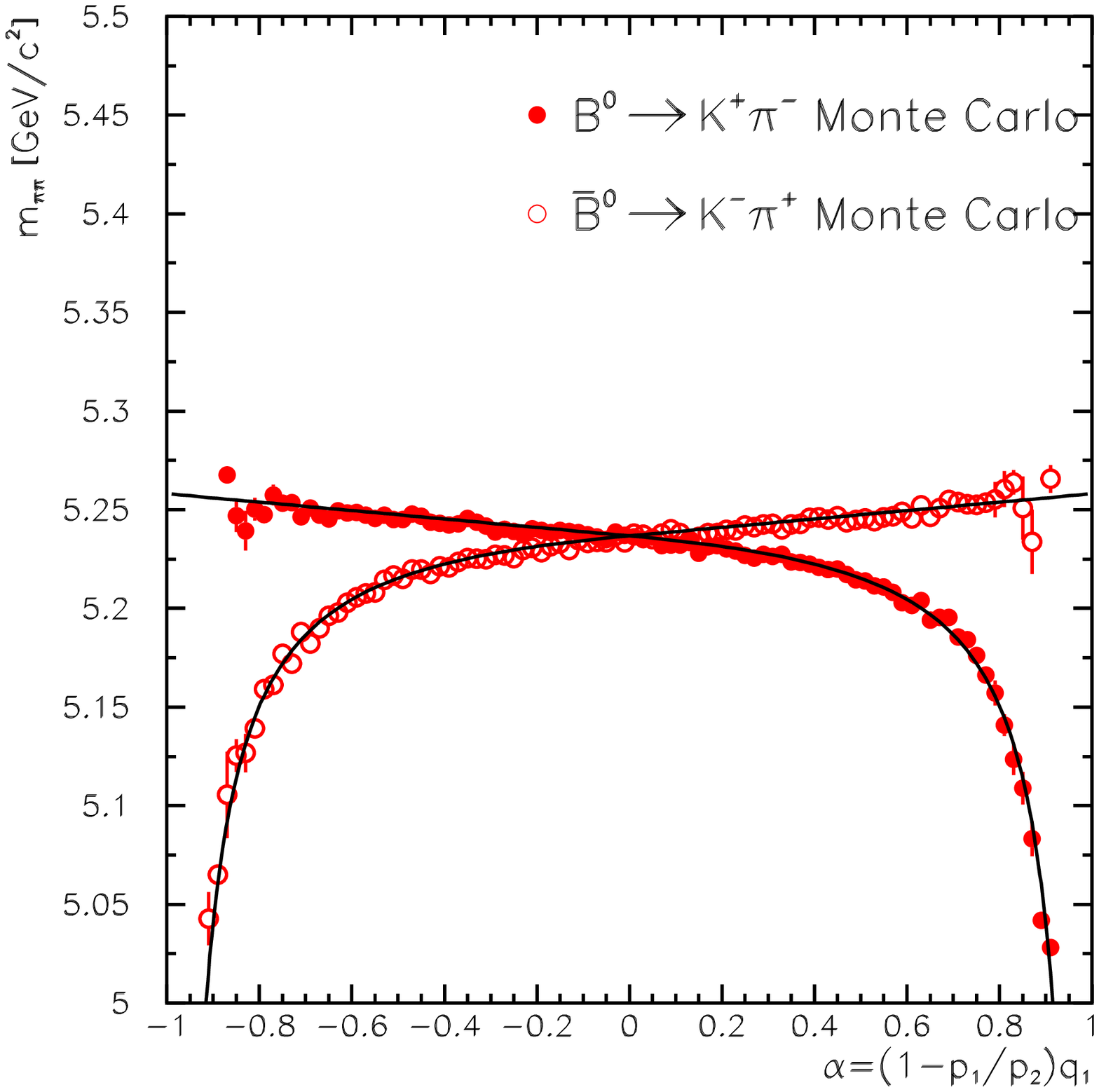}
\put(8,39){\scriptsize{\emph{(b)}}}								
\end{overpic}   
\begin{overpic}[width=4.69cm,height=4.69cm,grid=false,tics=1]{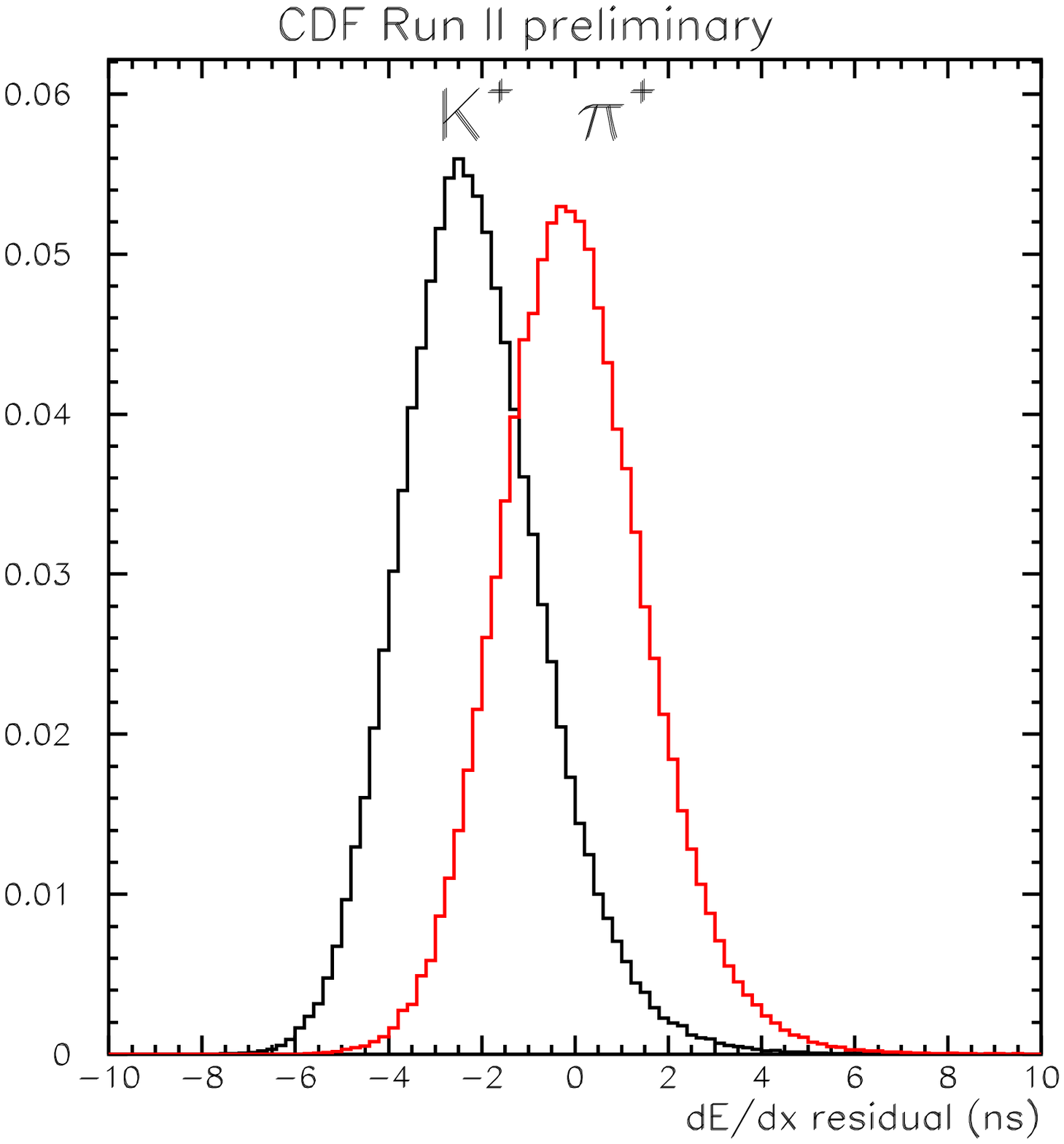}
\put(8,39){\scriptsize{\emph{(c)}}}								
\end{overpic}   
\end{center}
\caption{\emph{(a)}, $\pi\pi$-mass distribution 
after the optimized cuts; \emph{(b)},
 <$m_{\pi\pi}$> \emph{versus} $\alpha$ distribution for 
\bdkpi\ Monte Carlo events with the analytic function overlaid; \emph{(c)}, dE/dx residual
 for $K^+$ and $\pi^+$.}
\label{fig:plots}
\end{figure}

In the offline analysis, an unbiased procedure of optimization provided a
 tightened selection on track-pairs fit to a common vertex.  
 We found the optimal cuts by maximizing  the quantity $S/\sqrt{S+B}$.
 For each set of cuts, $S$ was the 
signal yield expected from Monte Carlo simulation, and $B$ were the background events found in
the sidebands of the $\pi\pi$-mass distribution in data.
Offline, we cut also on the $B$ meson \emph{isolation}:
  $I(B)= p_T(B)/[p_T(B) + \sum_{i} p_T(i)]$, 
in which the sum runs over every other track within a cone of radius 1  
in the $\eta-\phi$ space around the $B$ meson flight-direction.
This rejected 75\% of background while keeping 80\% of signal.
 The resulting $\pi\pi$-mass distribution  (Fig.~\ref{fig:plots}, (a)) shows a clean signal, estimated by a Gaussian (signal)
plus exponential (background) fit to contain $893 \pm 47$ events, with standard deviation $\sigma = 38 \pm 2$ MeV/$c^2$ and SNR$\approx$1.7 
at peak.
\section{Peak composition}
\label{sec:peak_composition}
Despite the excellent CDF II mass resolution\footnote{See some observed mass widths: $\sigma(J/\psi\rightarrow \mu^+\mu^-)\simeq 14$ \massmev, $\sigma(D^0\rightarrow K^+ \pi^-)\simeq 10$ \massmev.},
 the expected contributions to the signal ($B^0 \rightarrow K^+\pi^-$ and $\pi^+\pi^-$, $B^0_s \rightarrow K^+ K^-$ and $K^-\pi^+$)
 overlap into an unresolved peak, preventing an event-by-event
separation. We achieved a statistical separation instead, with an unbinned Likelihood fit that used PID information,
 provided by the dE/dx in the drift chamber, and kinematics.\par
We exploited the kinematic differences among modes by using an approximate relation between any
 two invariant masses $(M_{m_{1},m_{2}}$ and $M_{m_{1}',m_{2}'})$ obtained with two 
arbitrary mass-assignments to the tracks ($m_1, m_2$ and $m_1'$ $m_2'$).
 If $m_{1,2} \ll p_{1,2}$ we have~ $M^2_{m_1,m_2}\approx M^2_{m_1',m_2'}+\left(1+ p_1/p_2\right)\cdot\left(m_2^2 -m_2'^2\right)+\left(1+ p_2/p_1\right)\cdot\left(m_1^2 -m_1'^2\right)$, where the  kinematic information is contained in just two observables, 
a single candidate invariant-mass and the ratio of momenta.  The plot (b) in
 Fig.~\ref{fig:plots} shows  the averaged  $\pi\pi$-mass \emph{versus} the signed momentum-imbalance, $\alpha=(1-p_1/p_2)q_1$,
 for \bdkpi\ Monte Carlo events. The momentum (charge) $p_1$ ($q_1$) 
refers to the softer track; by combining kinematics and charge, we therefore separated also 
$B^0_{(s)}$ from $\bar{B}^0_{(s)}$ mesons in $K\pi$ modes.\par 
We equalized the dE/dx over the tracking volume and time using  
 $\sim$300,000  $D^{*+}\rightarrow D^0\pi^+\rightarrow [K^-\pi^+]\pi^+$ decays, 
where $D^0$ daughters were identified by the strong $D^{*+}$ 
decay\footnote{We neglected the $\sim 0.4\%$ 
contamination from  doubly Cabibbo-suppressed $D^0\rightarrow K^+\pi^-$ decays.} \cite{tesi}. In $>$95\% pure $K$ and $\pi$ samples,
we obtained 1.4$\sigma$ of $K/\pi$ separation (Fig.~\ref{fig:plots}, (c)).
We measured, and included in the fit, a  11\% residual track-to-track correlation due to 
common-mode dE/dx fluctuations.
\section{Fit and efficiency corrections}
\label{ssec:likelihood}
The fit used five observables: 
the $\pi\pi$-mass $m_{\pi\pi}$, the signed momentum-imbalance $\alpha$, the scalar sum 
of track's momenta, and the dE/dx of both tracks. 
The Likelihood for the single event $i$ was:
$\mathcal{L}^i = (1-b)\sum_{j=\mathit{mode}}f_j\mathcal{L}_j^{\mathit{kin}}\mathcal{L}_j^{\mathit{PID}} + b~\mathcal{L}_{\mathit{bck}}^{\mathit{kin}}\mathcal{L}_{\mathit{bck}}^{\mathit{PID}}$
where $j$ runs over the signal modes, and the output parameters $f_j$ ($b$) are the
 fractions of each mode (background). We obtained
the kinematic shape for signal  ($\mathcal{L}_j^{\mathit{kin}})$ partly
from the formula of Sec.~\ref{sec:peak_composition} and  partly
 from Monte Carlo. We fit the kinematic shape for background 
($\mathcal{L}_{\mathit{bck}}^{\mathit{kin}}$) in sideband events of Fig.~\ref{fig:plots}, (a).
We obtained the dE/dx shapes for signal ($\mathcal{L}_j^{\mathit{PID}}$)  and background 
($\mathcal{L}_{\mathit{bck}}^{\mathit{PID}}$) in pure $K$ and $\pi$
 from $D^0$ decays (Fig.~\ref{fig:plots}, (c)). 
\par The fit found three  contributions to the peak:
$121 \pm 27$ \bdpipi, $542 \pm 30$ \bdkpi, 
and $236 \pm 32$ \bskk\ decays. This is the first observation of the \bskk\ mode.\par
 We converted the fit results into ratios of branching fractions  by correcting for the differences in efficiencies
 between modes. We obtained part of the corrections 
from simulation (geometric acceptances, interaction and decay-in-flight probability of $K$ \emph{vs.} $\pi$, 
interaction probability of $K^+$ \emph{vs.} $K^-$), and part from samples of 
real data (dE/dx-dependent trigger efficiency and efficiency of the isolation cut).
The resulting total corrections varied between modes by less than 12\%.
\section{Results}
\label{sec:results}
 We observed the new decay \bskk. The measurement\footnote{In all results, the first error is statistical, 
the second systematic. $f_{d(s)}$ are the production fractions for $B^0_{(s)}$ mesons.} $(f_s/f_d)\cdot\mathcal{B}($\bskk$)/\mathcal{B}($\bdkpi$)= 0.46 \pm 0.08 \pm 0.07$\footnote{This result assumes
 the \bskk\ dominated by the short-lived component,
 $\Gamma_d = \Gamma_s$ and $\Delta\Gamma_s/\Gamma_s=0.12\pm 0.06$.} agrees with QCD-factorization predictions \cite{QCDF}, and it may 
 probe the size of U-spin violation,  possibly favoring the large SU(3)-breaking
 predicted by QCD sum-rules \cite{Khodjamirian}. For comparison with theory  \cite{matias},
 we also quote  $(f_d/f_s)\cdot\mathcal{B}($\bdpipi$)/\mathcal{B}($\bskk$)= 0.45 \pm 0.13 \pm 0.06$.  
We found no evidence of rarer \bshh\ modes, and we set 90\% C.L. upper limits \cite{FC} on their decay rates. These limits, 
$(f_s/f_d)\cdot\mathcal{B}($\bskpi$)/\mathcal{B}($\bdkpi$)<0.08$ and 
$\mathcal{B}($\bspipi$)/\mathcal{B}($\bskk$)<0.05$, are greatly improved over current world averages.\par In the $B^0$ sector, we 
 measured the direct \textsf{CP}-asymmetry of \bdkpi\ decay-rates:  $A_{\mathsf{CP}}= -0.013 \pm 0.078 \pm 0.012$,
in agreement with the results from the $B$-factories. With significantly more data already collected, the CDF measurement will soon
be competitive with their results.
We also quote $\mathcal{B}($\bdpipi$)/\mathcal{B}($\bdkpi$)=0.21 \pm 0.05 \pm 0.03$, used   
as a check of the validity of the whole analysis. We found
 no evidence of rarer \bdhh\ modes, and we set the following 90\% C.L. upper limit \cite{FC} on  the 
relative decay rate: $\mathcal{B}($\bdkk$)/\mathcal{B}($\bdkpi$)<0.10$.


\begin{thebibliography}{99}
\bibitem{fleischer} R.~Fleischer, \emph{New Strategies to Extract $\beta$ and $\gamma$ from 
$B_d\rightarrow \pi^+\pi^-$ and $B_s\rightarrow K^+K^-$}, Phys. Lett. B \textbf{459}, 306 (1999),
  [{\tt hep-ph/9903456}].
\bibitem{tdr} CDF II Collaboration, \emph{CDF II - Technical Design Report}, FERMILAB-PUB-96/390-E 
(1996).
\bibitem{SVT} W.~J.~Ashmanskas \emph{et al.}, \emph{The CDF Silicon Vertex Trigger}, Nucl. Instrum. Meth. A \textbf{518},
 532 (2004), [{\tt physics/0306169}]. 
\bibitem{tesi} D.~Tonelli, Ph.D. Thesis, Scuola Normale Superiore - Pisa, in preparation.
\bibitem{QCDF} M.~Beneke and M.~Neubert, \emph{QCD Factorization in $B \rightarrow PP$ and $B \rightarrow PV$ Decays}, Nucl. Phys. B \textbf{675}, 333 (2003), [{\tt hep-ph/0308039}].
\bibitem{Khodjamirian} A.~Khodjamirian \emph{et al.}, \emph{Flavour SU(3) Symmetry in  Charmless B Decays}, Phys. Rev. D \textbf{68}, 114007 (2003), [{\tt hep-ph/0308297}].
\bibitem{matias} D.~London and J.~Matias, \emph{Testing the Standard Model with  $B^0_{s}\rightarrow K^+ K^-$ Decays}, Phys. Rev. D \textbf{70}, 031502 (2004), [{\tt hep-ph/0404009}]. 
\bibitem{FC} G.~J.~Feldman and R.~D.~Cousins, \emph{A unified approach to the classical  statistical  analysis of small signals}, Phys. Rev. D \textbf{57}, 3873 (1998), [{\tt physics/9711021}].
\end{thebibliography}
\end{document}